\documentclass[12pt]{article}

\usepackage{epsf}
\usepackage{epsfig}
\usepackage{cite}
\usepackage{amsmath}
\usepackage{graphics}

\begin{document}

\setlength{\textheight}{21.5cm}
\setlength{\oddsidemargin}{0.cm}
\setlength{\evensidemargin}{0.cm}
\setlength{\topmargin}{0.cm}
\setlength{\footskip}{1cm}
\setlength{\arraycolsep}{2pt}

\renewcommand{\thefootnote}{\#\arabic{footnote}}
\setcounter{footnote}{0}

\newcommand{\gtrsim}{ \mathop{}_{\textstyle \sim}^{\textstyle >} }
\newcommand{\lesssim}{ \mathop{}_{\textstyle \sim}^{\textstyle <} }
\newcommand{\rem}[1]{{\bf #1}}
\renewcommand{\thefootnote}{\fnsymbol{footnote}}
\setcounter{footnote}{0}
\def\thefootnote{\fnsymbol{footnote}}

\hfill July 2009\\

\vskip .5in

\begin{center}

\bigskip
\bigskip

{\Large \bf $T^{'}$ Predictions of PMNS and CKM Angles}

\vskip .45in

{\bf Paul H. Frampton\footnote{frampton@physics.unc.edu}
and Shinya Matsuzaki\footnote{synya@physics.unc.edu}} 

\vskip .3in

{\it Department of Physics and Astronomy, University of North Carolina,
Chapel Hill, NC 27599-3255.}

\end{center}

\vskip .4in 
\begin{abstract}
Generalizing a previous model
to accommodate the third quark family and CP violation, we present 
a $T^{'}$ model which predicts tribimaximal neutrino (PMNS) mixings 
while the central predictions for
quark mixings are $|V_{td}/V_{ts}| = 0.245$ and $|V_{ub}/V_{cb}| = 0.237$
with a predicted CP violating KM phase $\delta_{KM} = 65.8^{0}$. All these 
are acceptably close to experiment,
including the KM phase for which the allowed values
are $63^{0} < \delta_{KM} < 72^{0}$, and depend only
on use of symmetry $T^{'} \times Z_2$
to define the model and no additional parameters.
\end{abstract}

\renewcommand{\thepage}{\arabic{page}}
\setcounter{page}{1}
\renewcommand{\thefootnote}{\#\arabic{footnote}}

\newpage

The justly celebrated standard model (SM) of particle 
phenomenology rests as one of the great accomplishments 
of theory.
Nevertheless, in the extended form which can accommodate
neutrino masses, it has 28 free parameters none of which
has been calculated or predicted. Notable
are quark and lepton masses (12, hereafter SM masses)
and mixing angles and phases (10, hereafter SM mixings)
which relate mass eigenstates for quarks and leptons
to states in the lagrangian. 

\bigskip

Except for a hint about the neutrino hierarchy, we find
no new insight about SM masses
\footnote{The result in an earlier paper \cite{FKM}
involving the quark mass squared ratio $(m_d^2/m_s^2)$
has only small corrections from the considerations
of the present Letter.}.
With regard to SM mixings we report at least limited progress
towards understanding 6 out of 10 phenomenological parameters
by the study of $T^{'}$ (binary tetrahedral group) as
a flavor symmetry commuting with the SM theory. If
even one of these explanations survives we feel it will
be useful and important.

\bigskip

In recent work, the present authors, together with
Kephart\cite{FKM}, presented a simplified model
based on $T^{'}$ flavor symmetry. The principal simplification
was that the CKM mixing angles 
\footnote{Note that here upper case $\Theta_{ij}$ refer to quarks (CKM) 
and lower case $\theta_{ij}$ will refer to neutrinos (PMNS).}
involving the third quark
family were taken to vanish $\Theta_{23} = \Theta_{13} = 0$.

\bigskip

In terms of the scalar field content, all scalar fields
are taken to be doublets under electroweak
$SU(2)$ with vacuum values which underly the symmetry breaking.
Great simplification was originally achieved by the device
of restricting scalar fields to irreducible representations 
of $T^{'}$ which are singlets and triplets only, without any $T^{'}$ doublets.
There was a good reason for this because the admission of
$T^{'}$-doublet scalars enormously complicates the symmetry
breaking.  This enabled the isolation of the Cabibbo angle $\Theta_{12}$
and to a reasonable prediction thereof, namely
\cite{FKM} $\tan 2 \Theta_{12} = (\sqrt{2})/3$.

\bigskip

Within the same simplified model, in a subsequent paper with Eby \cite{EFM}, 
the departure of $\Theta_{12}$ from this
$T^{'}$ prediction was used to make predictions for the departure
of the neutrino PMNS angles $\theta_{ij}$ from their
tribimaximal values\cite{HPS}. 
Also in that model \cite{FM}, we
suggested a smoking-gun $T^{'}$ prediction
for leptonic decay of the standard model Higgs scalar.
Other related works are 
\cite{FMA4,Ma,Ma2,Altarelli,Kephart0,Feruglio,mahanthappa,HS,textures,marfatia,FGY}.

\bigskip

In the present article, we examine the addition
of $T^{'}$-doublet scalars. As anticipated in \cite{FKM}, this
allows more possibilities of $T^{'}$ symmetry breaking
and permits non-zero values for $\Theta_{23}$, $\Theta_{13}$
and $\delta_{KM}$.  
We present an explicit $(T^{'} \times Z_2)$
model which leads to 
consistent results for PMNS and KM angles
as well as testable predictions.

\bigskip

To understand the incorporation of $T^{'}$-doublet
scalars and to make the present article self-contained, 
it is necessary  to review the simplified
model employed in \cite{FKM,EFM,FM} in which $T^{'}$-doublet
scalars were deliberately excluded in order to isolate
the Cabibbo angle $\Theta_{12}$.
We here adopt the global symmetry $(T^{'} \times Z_2)$.

\bigskip
\bigskip

Left-handed quark doublets \noindent $(t, b)_L, (c, d)_L, (u, d)_L$
are assigned under this as 

\begin{equation}
\begin{array}{cc}
\left( \begin{array}{c} t \\ b \end{array} \right)_{L}
~ {\cal Q}_L ~~~~~~~~~~~ ({\bf 1_1}, +1)   \\
\left. \begin{array}{c} \left( \begin{array}{c} c \\ s \end{array} \right)_{L}
\\
\left( \begin{array}{c} u \\ d \end{array} \right)_{L}  \end{array} \right\}
Q_L ~~~~~~~~ ({\bf 2_1}, +1),
\end{array}
\label{qL}
\end{equation}

\noindent and the six right-handed quarks as

\begin{equation}
\begin{array}{c}
t_{R} ~~~~~~~~~~~~~~ ({\bf 1_1}, +1)   \\
b_{R} ~~~~~~~~~~~~~~ ({\bf 1_2}, -1)  \\
\left. \begin{array}{c} c_{R} \\ u_{R} \end{array} \right\}
{\cal C}_R ~~~~~~~~ ({\bf 2_3}, -1)\\
\left. \begin{array}{c} s_{R} \\ d_{R} \end{array} \right\}
{\cal S}_R ~~~~~~~~ ({\bf 2_2}, +1).
\end{array}
\label{qR}
\end{equation}

\bigskip

\noindent The leptons are assigned as
\begin{equation}
\begin{array}{ccc}
\left. \begin{array}{c}
\left( \begin{array}{c} \nu_{\tau} \\ \tau^- \end{array} \right)_{L} \\
\left( \begin{array}{c} \nu_{\mu} \\ \mu^- \end{array} \right)_{L} \\
\left( \begin{array}{c} \nu_e \\ e^- \end{array} \right)_{L}
\end{array} \right\}
L_L  (3, +1)  &
\begin{array}{c}
~ \tau^-_{R}~ (1_1, -1)   \\
~ \mu^-_{R} ~ (1_2, -1) \\
~ e^-_{R} ~ (1_3, -1)  \end{array}
&
\begin{array}{c}
~ N^{(1)}_{R} ~ (1_1, +1) \\
~ N^{(2)}_R ~ (1_2, +1) \\
~ N^{(3)}_{R} ~ (1_3, +1),\\  \end{array}
\end{array}
\end{equation}

\bigskip

Next we turn to the symmetry breaking and the necessary
scalar sector with its own potential
\footnote{The scalar potential will not be examined
explicitly. We assume that it has enough parameters 
to accommodate the required VEVs in a finite neighborhood of 
parameter values.}
and Yukawa coupling to the fermions, leptons and quarks.

\bigskip
\bigskip

For scalar fields we keep those used previously 
namely the two $T^{'}$ triplets and two $T^{'}$ singlets
\begin{equation}
H_3(3, +1); ~ H_3^{'}(3, -1); ~ H_{1_1}(1_1, +1); ~ H_{1_3}(1_3, -1)
\label{oldscalars}
\end{equation}
which led to the simplified model discussed previously
with CKM angles $\Theta_{23} = \Theta_{13} = 0$. That
model was used to derive a formula for
the Cabibbo angle\cite{FKM}, to predict corrections\cite{EFM}
to the tribimaximal values\cite{HPS} of PMNS neutrino
angles, and to make a prediction for
Higgs boson decay\cite{FM}.

\bigskip
\bigskip

We now introduce one $T^{'}$ doublet scalar in an explicit model. 
Non-vanishing $\Theta_{23}$ and $\Theta_{13}$ will be induced by
symmetry breaking due to the addition
the $T^{'}$ doublet scalar
\begin{equation}
H_{2_3}(2_3, +1)
\label{newscalar}
\end{equation}

\bigskip

\noindent The field in Eq.(\ref{newscalar}) allows a new
\footnote{We list the Yukawa couplings already discussed
for the $T^{'}$-triplet and $T^{'}$-singlet scalars:

\begin{eqnarray}
{\cal L}_Y
&=&
\frac{1}{2} M_1 N_R^{(1)} N_R^{(1)} + M_{23} N_R^{(2)} N_R^{(3)} \nonumber \\
& & + \Bigg\{
Y_{1} \left( L_L N_R^{(1)} H_3 \right) + Y_{2} \left(  L_L N_R^{(2)}  H_3
\right) + Y_{3}
\left( L_L N_R^{(3)} H_3 \right)  \nonumber \\
&& +
Y_\tau \left( L_L \tau_R H'_3 \right)
+ Y_\mu  \left( L_L \mu_R  H'_3 \right) +
Y_e \left( L_L e_R H'_3 \right)
\Bigg\} \nonumber \\
&& + Y_t ( \{{\cal Q}_L\}_{\bf 1_1}  \{t_R\}_{\bf 1_1} H_{\bf 1_1}) \nonumber \\
&&
+ Y_b (\{{\cal Q}_L\}_{\bf 1_1} \{b_R\}_{\bf 1_2} H_{\bf 1_3} ) \nonumber \\
&&
+ Y_{{\cal C}} ( \{ Q_L \}_{\bf 2_1} \{ {\cal C}_R \}_{\bf 2_3} H^{'}_{\bf 3})
\nonumber \\
&&
+ Y_{{\cal S}} ( \{ Q_L \}_{\bf 2_1} \{ {\cal S}_R \}_{\bf 2_2} H_{\bf 3})
\nonumber \\
&&
+ {\rm h.c.}. \nonumber
\end{eqnarray}
}
Yukawa coupling

\begin{equation}
Y_{{\cal Q}{\cal S}} {\cal Q}_L {\cal S}_R H_{2_3} + h.c.
\label{newYukawa}
\end{equation}
where $Y_{sb} = Y_{sb}^{*}$ is real and we
accommodate CP violation by the use of complex $T^{'}$ Clebsch-Gordan
coefficients.

\bigskip

\noindent The vacuum expectation value (VEV) for $H_{2_3}$ is
taken with the alignment

\begin{equation}
<H_{2_3}> = V_{2_3} (1, 1)
\label{vev}
\end{equation}
while as in \cite{FKM} the other VEVs include
\begin{equation}
<H_3> = V(1, -2, 1)
\label{H3vev}
\end{equation}

\bigskip

The hermitian squared mass matrix ${\cal D} \equiv D D^{\dagger}$ for the charge
$(-1/3)$ quarks is then

\begin{equation}
{\cal D} =
\left( \begin{array}{ccc}
m_b^2 & \frac{1}{\sqrt{6}} Y_{{\cal S}} Y_{{\cal Q}{\cal S}} V V_{2_3} (1-2\sqrt{2}\omega^2) &   
\frac{1}{\sqrt{6}} Y_{{\cal S}} Y_{{\cal Q}{\cal S}} V V_{2_3} (\omega^2 + \sqrt{2}) \\  
\frac{1}{\sqrt{6}} Y_{{\cal S}} Y_{{\cal Q}{\cal S}} V V_{2_3} (1 - 2\sqrt{2}\omega^{-2}) &   
3 (Y_{{\cal S}} V)^2 & -\frac{\sqrt{2}}{3} (Y_{{\cal S}} V)^2 \\
\frac{1}{\sqrt{6}} Y_{{\cal S}} Y_{{\cal Q}{\cal S}} V V_{2_3} (\omega^{-2} + \sqrt{2})   
&
-\frac{\sqrt{2}}{3} (Y_{{\cal S}} V)^2 &  (Y_{{\cal S}} V)^2 
\end{array}
\right)
\label{calD}
\end{equation}

\bigskip

\noindent In Eq.(\ref{calD}) the $2\times2$ sub-matrix for the first two families
coincides with the result discussed earlier \cite{FKM} and hence the successful
Cabibbo angle formula $\tan 2\Theta_{12} = (\sqrt{3})/2$ is preserved.

\bigskip

\noindent Note that in this model the mass matrix for the charge $+2/3$ quarks is diagonal
\footnote{This uses the approximation that the electron mass is $m_e=0$;
{\it c.f.} ref.\cite{FKM}.}
so the CKM mixing matrix arises purely from diagonalization of ${\cal D}$ in Eq.(\ref{calD}).
The presence of the complex $T^{'}$ Clebsch-Gordan
coefficients \cite{Chen} in Eq.(\ref{calD}) which will lead to non-zero
KM CP violating phase.

\bigskip

\noindent For $m_b^2$ the experimental value $17.6 GeV^2$ \cite{PDG2008}
although the CKM angles and phase do not depend on this
overall normalization.

\bigskip

\noindent Actually our results depend only on assuming that the ratio
$(Y_{{\cal Q}{\cal S}} V_{2_3}/ Y_{{\cal S}} V)$ is much smaller than one.

\bigskip

\noindent In this case we find the results for the ratios of CKM matrix elements

\begin{equation}
|V_{td}/V_{ts}| = 0.245
\end{equation}

\noindent and 

\begin{equation}
|V_{ub}/V_{cb}| = 0.237
\end{equation}

\bigskip
\bigskip

\noindent For the CP violating phase $\delta_{KM}$ 

\begin{equation}
\delta_{KM} = 65.8^{0}
\label{KMphase}
\end{equation}
is in the experimentally allowed range
$63^{0} < \delta_{KM} < 72^{0}$. 

\bigskip

\noindent More technical details will be provided
in \cite{EFM2}.

\bigskip

\noindent Note that once the off-diagonal
third-family elements
in Eq.(\ref{calD}) are taken as much smaller
than the elements involved in the
Cabibbo angle, the two KM angles and the CP phase
are predicted without further free parameters
so this vindicates the hope expressed in \cite{FKM}.

\bigskip

\noindent In summary, we have reported progress in understanding the PMNS and CKM
mixing angles by exploiting the binary tetrahedral group ($T^{'}$)
as a global discrete flavor symmetry commuting with the 
local gauge symmetry $SU(3) \times SU(2) \times U(1)$ of
the standard model of particle phenomenology. The results
are encouraging to pursue this direction of study.

\newpage

\begin{center}

\section*{Acknowledgements}

\end{center}

This work was supported in part 
by the U.S. Department of Energy under Grant
No. DE-FG02-06ER41418.

\bigskip
\bigskip
\bigskip
\bigskip

\end{document}